\documentclass[12pt]{article}
\usepackage{graphicx}
\usepackage{epsfig}
\begin{document}
\vspace*{2cm} \noindent {\Large\bf A Droplet within the Spherical
Model.} \vspace{\baselineskip}
\newline {\bf A.E.\ Patrick\footnote[1]{Laboratory of Theoretical
Physics, Joint Institute for Nuclear Research, Dubna 141980,
Russia e-mail: patrick@theor.jinr.ru}}
\begin{list}{}{\setlength{\rightmargin }{0 mm}
\setlength{\leftmargin }{2.5cm}} \item \rule{124mm}{0.3mm}
\linebreak {\footnotesize {\bf Abstract.} Various substances in the liquid state
tend to form droplets. In this paper the shape of such droplets is investigated
within the spherical model of a lattice gas. We show that in this case the droplet
boundary is always diffusive, as opposed to sharp, and find the corresponding density
profiles (droplet shapes). Translation-invariant versions of the spherical model do
not fix the spatial location of the droplet, hence lead to mixed phases.
To obtain pure macroscopic states (which describe localized droplets) we use
generalized quasi-averaging. Conventional quasi-averaging deforms droplets
and, hence, can not be used for this purpose. On the contrary, application of
the generalized method of quasi-averages yields droplet shapes which do not
depend on the magnitude of the applied external field.
}
\newline \rule[1ex]{12.4cm}{0.3mm}
\linebreak {\bf \sc key words:} {\footnotesize Droplet shape;
lattice gas; pure phases; quasi-averages.}
\vspace{\baselineskip}
\end{list}

\section{Introduction.} The purpose of the Gibbs distribution and
the statistical physics in general is to provide a bridge between
microscopic and macroscopic phenomena. It would be a mistake
though to think that the bridge is in any sense similar to one of,
for instance, London bridges. The difficulties of actual
``traveling" through that bridge in the case of
finite-dimensional models (with $d\geq 2$) are so formidable, that
a better name for the bridge might have been a labyrinth.
Therefore it was not really surprising that it were mathematicians
(not physicists) who actually managed to walk through a few
bridges connecting microscopic interparticle interactions with
macroscopic shapes of droplets of condensed matter.

In a colossal effort beginning from works of Minlos and Sinai
\cite{ms67,ms68} it was shown that for sufficiently low
temperatures typical configurations of discrete lattice models
look like a macroscopic droplet of one phase surrounded
by another phase. The droplet border is sharp in the macroscopic
scale, that is, the magnitude of typical fluctuations of the
corresponding long contour is much smaller than the linear size of
the droplet. Somewhat later the exact shape of the droplet in the
case of the Ising model was also found. It is given by a curve
minimizing the corresponding Wulff functional. For an accurate
account of all ``twists and turns", which should be dealt with in
order to walk through the bridge-labyrinth connecting the shape
of a macroscopic droplet with the interaction of Ising particles,
see the books \cite{dks92,s91}.

The behaviour of continuous lattice systems is different from that
of the above mentioned discrete systems, and properties of Gibbs
states of, for instance, $O(n)$-models are much less understood
than those of various discrete models. Fortunately, there is a
continuous lattice model which macroscopic properties can be
derived from the interaction of microscopic variables with only
modest efforts. This model is the, so-called, spherical lattice
gas \cite{gb62,pk60}.

The authors of the paper \cite{pk60} studied the equation of state
of the model --- the behaviour of pressure as a function of
specific volume and temperature. At the time, derivation of
equation of state from a microscopic interaction was an
achievement on its own right. Therefore, it is not surprising that
such a question as the shape of a droplet of condensed ``spherical matter"
was not even considered. Moreover, the spherical lattice gas with
cyclic boundary conditions is a translation invariant model.
Hence, the center of the droplet is uniformly distributed over the
available volume, and the constant average values of microscopic
variables do not reveal the droplet shape.

At nearly the same time, in the paper \cite{b60} similar phenomena
related to invariance of correlation functions were termed the
degeneracy of equilibrium state, and a heuristic procedure to
remedy the situation --- the quasi-averages --- was advertised.
Calculation of quasi-averages involves switching on an appropriate
symmetry-breaking field of a magnitude proportional to
$\varepsilon$, and switching off the field by sending
$\varepsilon\downarrow 0$ after the thermodynamic limit. It is
shown in the present paper that the conventional quasi-averages
allow one to find the true shape of the droplet only in a fortuitous
situation when the symmetry-breaking field is already similar to
the droplet shape we are looking for. The field of a different
shape not only fixes the location of the droplet but also deforms
it unrecognizably. Nevertheless, one can say that certain tools
for finding the shape of a droplet in the spherical lattice gas
were available already in early sixties, but investigation of that
kind can not be found in the literature.

Two decades later the dominant terminology became
pure phases and mixed phases (instead of non-degenerate and
degenerate equilibrium states, respectively), see, e.g., the book
\cite{gj81}. More importantly, a simple criterion allowing one to
answer the question whether a phase is pure or mixed became widely
known: a phase is pure if the covariance of microscopic dynamic variables
associated with nodes $j$ and $k$ tend to zero as the distance $|j-k|$
increases. Pure phases can be obtained with a help of quasi-averages,
or by using appropriate boundary conditions, or by calculating
conditional distributions. Mixed phases can always be represented
as linear combinations of pure phases.

Gersch and Berlin, see \cite{gb62}, found the (constant) expected
values, covariances, and conditional expected values  of
microscopic variables of the spherical lattice gas. The derived
covariances do not tend to 0 with the distance between the corresponding
microscopic degrees of freedom, hence, the natural
phase of the lattice gas is not pure.  In order to obtain the
droplet shape in a situation like that, it is necessary to
decompose the mixed phase into pure components. However,
apparently, the authors of the paper \cite{gb62} did not realize
that.

In the present paper we extract pure phases of the spherical
lattice gas  using (generalized) quasi-averages, see \cite{az85,bzt86}.
The investigation of the properties of the pure phases shows that the
droplet in the spherical lattice gas is always diffusive. That is,
the boundaries of the droplet are not sharp, not even in the macroscopic
scale. The constant levels of the expected values of microscopic variables
look like rounded squares, although not exactly the same ones as the
rounded squares describing the sharp boundaries of the droplet within
the 2D Ising model of a lattice gas.


The rest of the paper is organized as follows. Section 2 contains
a precise definition of the spherical lattice gas. It also
contains some well known technical results for the use in the
later sections. Section 3 introduces macro states and summarizes
the main results of the paper. In Section 4 we calculate the
distributions of microscopic
random variables in the mixed phase of the spherical lattice gas.
In Section 5 we use the Lagrange method to find pure
macroscopic phases for the lattice gas with periodic boundary
conditions at zero temperature. Section 6 is the main part of this paper.
There we use the method of quasi-averages for extracting
pure macroscopic phases.
The results of the paper are discussed in Section 7.

\section{The model and useful facts.}
The spherical lattice gas is a collection of random variables
$\{x_j,j\in Z^d\}$ placed at sites of an integer
$d$-dimensional lattice, $Z^d$. Every site $j\in Z^d$ is specified
by its $d$ integer coordinates $(j_1,j_2,...,j_d)$. In the present
paper we consider the case $d\geq3$.

To define the distribution of random variables at all sites of the
lattice, we first specify the joint distribution for the random
variables in a finite cube
\[
V_n=\{j\in Z^d:1\leq j_\nu\leq n,\nu=1,2,...,d\/\}
\]
containing $N\equiv n^d$ sites, and then pass to the limit
$n\to\infty$. To avoid unnecessary complications we impose
periodic boundary conditions in all dimensions.

It is instructive to consider also a stretched model defined on
the parallelepipeds
\begin{equation}
\Upsilon_n=\{j\in Z^d:1\leq j_\nu\leq n_\nu,\nu=1,2,...,d\/\},
\label{stretch}
\end{equation}
where $n_1=(1+\delta) n$, $\delta>0$, and $n_\nu=n$, for
$\nu=2,3,...,d$. Save for one side being longer than the others,
the definition of the stretched model is exactly the same as that
of the conventional spherical lattice gas.
\vspace{\abovedisplayskip}

\underline{\bf The Hamiltonian.} \vspace{\abovedisplayskip}

\noindent The random variables located in the rectangle $V_n$
interact with each other and with the external field $\{h_{j},j\in
Z^d\}$ via the Hamiltonian
\begin{equation}
H_n=-J\sum_{j,k\in V_n}T_{jk}x_j x_k-\sum_{j\in V_n}h_{j}x_j,
\label{ham}
\end{equation}
where $J>0$, and $T_{jk}$ are the elements of the
nearest-neighbour interaction matrix. In this paper the field
$\{h_{j},j\in Z^d\}$ is used as a technical tool, so that, it
should not necessarily be physically sensible. We let its
magnitude to depend on the size of rectangle $V_n$:
\begin{equation}
h_j=n^{-\gamma} b_j,\quad j\in Z^d, \label{ft}
\end{equation}
where the absolute values of $b_j$ are bounded by an independent of
$n$ constant $b$.

\vspace{\abovedisplayskip}

\underline{\bf The interaction matrix.} \vspace{\abovedisplayskip}

\noindent The elements of the interaction matrix $\widehat{T}$ describe the
usual nearest neighbour interaction on a square lattice, and they are
given by
\[
T_{jk}=\sum_{\nu=1}^d
J^{(\nu)}(j_\nu,k_\nu)\prod_{l\in\{1,2,\ldots,d\}\setminus\nu}\delta(j_l,k_l),
\]
where
\[
\delta(j_l,k_l)=\left\{
\begin{array}{cl}1,&\mbox{ if }j_l=k_l,\\
0,&\mbox{ if }j_l\neq k_l,
\end{array}
\right.
\]
is the Kronecker delta.

The coefficients $J^{(\nu)}(j_\nu,k_\nu)$, for $\nu=1,2,\ldots,d$,
are the elements of the matrix
\[
\widehat{J}=\left(
\begin{array}{ccccccc}0&\frac{1}{2}&&&&&\frac{1}{2}\\
\frac{1}{2}&0&\frac{1}{2}&&&\mbox{\LARGE 0}&\\
&\frac{1}{2}&0&\ddots&&&\\
&&\ddots&\ddots&\ddots&&\\
&&&\ddots&0&\frac{1}{2}&\\
&\mbox{\LARGE 0}&&&\frac{1}{2}&0&\frac{1}{2}\\
\frac{1}{2}&&&&&\frac{1}{2}&0
\end{array}
\right).
\]
The eigenvalues and orthonormal eigenvectors of the matrix
$\widehat{J}$ are given by
\[
\lambda_l=\cos\frac{2\pi (l-1)}{n},\quad l=1,2,\ldots,n,
\]
and
\[
\mbox{\boldmath $u$}^{(l)}
=\left\{u_m^{(l)}=\sqrt{\frac{2}{n}}\cos\left[\frac{2\pi (l-1)
(m-1)}{n}-\frac{\pi}{4}\right]\right\}_{m=1}^n,\quad
l=1,2,\ldots,n.
\]

The eigenvalues of the interaction matrix $\widehat{T}$ are the
sums of eigenvalues of the matrix $\widehat{J}$:
\[
\Lambda_k=\sum_{\nu=1}^d\lambda_{k_\nu},\quad
k\equiv(k_1,k_2,\ldots,k_d)\in V_n.
\]
The corresponding orthonormal eigenvectors are the products of
eigenvectors of the matrix $\widehat{J}$:
\begin{equation}
\mbox{\boldmath $w$}^{(k)} =\left\{w_j^{(k)}=\prod_{\nu=1}^d
u_{j_\nu}^{(k_\nu)}\right\}_{j\in V_n},\quad
k\equiv(k_1,k_2,\ldots,k_d)\in V_n. \label{evs}
\end{equation}

Note that the second-largest eigenvalue $\Lambda_{\,\rm s}\equiv
d-1+\cos(2\pi/n)$ of the interaction matrix (which will play an
important role below) is $2d$ times degenerate. At the same time,
in the case of the stretched  model the second-largest eigenvalue
$\Lambda_{\,\rm s}\equiv
d-1+\cos\left[\frac{2\pi}{(1+\delta)n}\right]$ is only twice
degenerate. \vspace{\abovedisplayskip}

\underline{\bf The Gibbs distribution.} \vspace{\abovedisplayskip}

The joint distribution of the random variables $\{x_j,j\in V_n\}$
is specified by the usual Gibbs density
\[
p(\{x_j,j\in V_n\})=\frac{e^{-\beta H_n}}{\Theta_n(\rho)},
\]
with respect to the {\em ``a priori''} measure
\[
\mu_n(dx)=\delta\left(\sum_{j\in V_n}x_j-\rho
N\right)\delta\left(\sum_{j\in V_n}x_j^2-N\right)\prod_{j\in
V_n}dx_j.
\]
The first delta function imposes the typical for gas models density constraint
\[
\frac{1}{N}\sum_{j\in V_n}x_j=\rho,
\]
the second one imposes the usual spherical constraint
\[
\sum_{j\in V_n}x_j^2=N.
\]

The normalization factor (partition function) $\Theta_n(\rho)$ is given
by
\begin{equation}
\Theta_n(\rho)=\int_{-\infty}^\infty\ldots\int_{-\infty}^\infty
e^{-\beta H_n}\mu_n(dx). \label{pf}
\end{equation}

\section{Macro states and the main results.}

The usual Gibbs states provide a detailed {\em microscopic\/}
description of a thermodynamic system. In some sense the amount of
available detail is too big: there can be a macroscopically
inhomogeneous structure in the thermodynamic system, but its shape
can not be described within the Gibbs-state framework. In this
situation an introduction of a rougher (reduced) description seems
justified and useful.

We define {\em macro states\/} as the following continuum limit of
the original lattice system. The limiting configurations are
realizations of random functions defined on the $d$-dimensional
rectangle $[0,1]^d$:
\[
\{x(\gamma)\}_{\gamma\in[0,1]^d}\equiv\{x(\gamma_1,\gamma_2,\ldots,
\gamma_d)\}_{\gamma_1,\gamma_2,\ldots,\gamma_d\in[0,1]}.
\]
For any $\gamma\in[0,1]^d$ the random variable $x(\gamma)$ is
defined as the following limit in distribution
\[
x(\gamma)\stackrel{d}{=}\lim_{n\to\infty}x_{([\gamma_1n],[\gamma_2n],\ldots,[\gamma_dn])},
\]
where $[y]$, is the integer part of $y$. Of course, for correctness of the above definition we need some
kind of continuity in the system, so that for two sequences $j(n)$
and $k(n)$ with the same limits
\[
\lim_{n\to\infty}(j_1(n)/n,\ldots,j_d(n)/n)=
\lim_{n\to\infty}(k_1(n)/n,\ldots,k_d(n)/n)=
(\gamma_1,\gamma_2,\ldots,\gamma_d),
\]
we also have identical limits of the corresponding sequences of
random variables $\{x_{j(n)}\}_{n=1}^\infty$ and
$\{x_{k(n)}\}_{n=1}^\infty$. Although in discrete spin systems
that kind of continuity may well be missing, we do not worry too
much about that, because it should be possible to surpass this
technical problem one way or another.

Thermodynamic random variables $x(\gamma)$ and $x(\delta)$ are
limits of the random sequences
$x_{([\gamma_1n],[\gamma_2n],\ldots,[\gamma_dn])}$ and
$x_{([\delta_1n],[\delta_2n],\ldots,[\delta_dn])}$ separated by a
distance of order $n$. Hence, in the continuum limit the random
variables $x(\gamma)$ and $x(\delta)$ with $\gamma\neq\delta$ are
independent due to the exponential/power-law decay of correlations
in pure phases of high/low temperature regions. Therefore, a
pure macro state is completely characterized by individual
distributions
\[
F_\gamma(y)=\lim_{n\to\infty}\mbox{
Pr}_{\,n}\left[x_{([\gamma_1n],[\gamma_2n],\ldots,[\gamma_dn])}\leq
y\right].
\]

Macro states defined above contain complete information about the
microscopic individual distributions of the random variables
$x_j$. It might be desirable to achieve further reduction of
description by using, for instance, Kadanoff's blocks in the
definition of the continuum limit. That should reduce the range of
limiting distribution outside the critical lines/points to just
the normal distribution. Even on critical lines the asymptotic
distributions of large Kadanoff's blocks should be of quite
limited variety.\vspace{\abovedisplayskip}

The main results of the present paper obtained for the
low-temperature region $\beta>\beta_c(\rho)$, see Eq.\ (\ref{ct}),
can be stated as follows.

\begin{enumerate}
\item The expected values of the random variables $x_j$ in the
natural state of the spherical lattice gas are translation
invariant and equal to the density, $\langle x_j\rangle=\rho$.
Their variances and covariances in the limit $n\to\infty$ are
given by
\[
\mbox{Var}(x_j)=\frac{W_d(d)}{2\beta
J}+\frac{1-\rho^2}{2}\left(1-\frac{\beta_c}{\beta}\right)+o(1),
\]
\[
\mbox{Cov}(x_j,x_l)= c(j,l)+
(1-\rho^2)\left(1-\frac{\beta_c}{\beta}\right)
\frac{1}{d}\sum_{\nu=1}^d\cos\frac{2\pi(j_\nu-l_\nu)}{n}+o(1),
\]
where $c(j,l)$ are the covariances of the microscopic variables in
the ordinary spherical model, see Eq. (\ref{ucf}).

\item The random variables $x_j$ within the stretched model admit
the following representation
\[
x_j=\sqrt{2(1-\rho^2)\left(1-\frac{\beta_c}{\beta}\right)}\sin(2\pi\!\mbox{
\boldmath $U$})+{\cal N}_j\left(\rho,\frac{W_d(d)}{2\beta
J}\right),
\]
where the random variable {\boldmath $U$} is uniformly distributed
on the interval $[0,1]$, ${\cal N}_j(a,b)$ are normal random variables
with mean $a$ and variance $b$, and
\[
\mbox{Cov}({\cal N}_j,{\cal N}_l)\equiv c(j,l)
\to 0\ \mbox{as dist}(j,l)\to\infty.
\]

\item For $T=0$ the pure equilibrium states of the model (ground
states) are given by
\[
x_j=\rho+\sqrt{2(1-\rho^2)}\sum_{\nu=1}^d
r_\nu\cos\left[\frac{2\pi(j_\nu-1)}{n}-\alpha_\nu\right],
\]
where $\alpha_\nu$ and $r_\nu$ are any constants
satisfying $\alpha_\nu\in[0,2\pi]$ and $\sum_{\nu=1}^d r^2_\nu = 1$.

\item In the presence of a symmetry-breaking field
\[
h_j=\left\{
\begin{array}{cl}
\varepsilon n^{-\delta},&\mbox{ if }j_1=1, \\
0,&\mbox{ otherwise,}
\end{array}
\right.
\]
the expected values of the random variables $x_j$ are given by
\[
\langle x_j\rangle\sim \frac{\varepsilon
n^{(1-\delta)/3}}{2J\sqrt{2\zeta^*}}
\left[1+\sqrt{2\zeta^*}n^{-(1+2\delta)/3}\right]^{-j_1},
\]
if $0<\delta<1$ and $j_1\leq n/2$. In this case all condensed
``spherical matter" gathers in a narrow strip around the plane
where the field is applied. The shape of the droplet is clearly
deformed by the field.

If $\delta=1$ the distributions of the random variables $x_j$
still depend on the magnitude of the symmetry breaking field, but
in the limit $n\to\infty$ they remain proper. In particular, the
finite expected values of the random variables $x_j$ are given by Eq.\
(\ref{evs2}). The droplet shape is still deformed by the field.

If $1<\delta<d-1$, then the limiting distributions of the random
variables $x_j$ do not depend on the magnitude of the symmetry
breaking field. In particular, the expected values of the random
variables $x_j$ are given by
\[
\langle x_j\rangle=\rho+\sqrt{2(1-\rho^2)
\left(1-\frac{\beta_c}{\beta}\right)}\cos(2\pi\gamma_1)+o(1).
\]
The droplet shape is determined by the field (for different field types
we obtain different shapes), but it is no longer deformed by the field
(does not depend on $varepsilon$).

%
%
%
\end{enumerate}

\section{The mixed phase.}

Main thermodynamic properties of the spherical lattice gas with
{\em cyclic\/} boundary conditions were derived in the papers
\cite{gb62,pk60}. Although the free energy density of the model is
not sensitive to the type of boundary conditions used,
more delicate properties, like the droplet shape, are. Therefore in this section
we re-derive the results of the paper \cite{gb62} in the case of,
more realistic, {\em periodic\/} boundary conditions.

The calculation of free energy, expected values, and correlation
functions for the spherical lattice gas is reduced, in a routine
fashion, to calculation of the large-$n$ asymptotics of an
integral, see \cite{gb62}.
Introduction of new integration variables $y_j$, $j\in V_n$ in
Eq.\ (\ref{pf}) via the orthogonal transformation
\begin{equation}
x_j=\sum_{k\in V_n}w_j^{(k)}y_k,\quad j\in V_n,
\label{nc}
\end{equation}
where the eigenvectors $\{w_j^{(k)},j\in V_n\}$ are given by Eq.\
(\ref{evs}), diagonalises the interaction matrix. As a result, we
obtain the following expression for the partition function
\[
\Theta_n(\rho)=\int_{-\infty}^\infty\ldots\int_{-\infty}^\infty
\exp\left(\beta J\sum_{k\in V_n}\lambda_k y_k^2+\beta\sum_{k\in
V_n}\varphi_k y_k\right) \widetilde\mu_n(dy),
\]
where $\varphi_k=\sum_{j\in V_n}h_jw_j^{(k)}$, and
\[
\widetilde\mu_n(dy)=\delta\left(y_{(1,1,\ldots,1)}\sqrt{N}-\rho
N\right)\delta\left(\sum_{j\in V_n}y_j^2-N\right)\prod_{j\in
V_n}dy_j.
\]
Integration over $y_{(1,1,\ldots,1)}$ yields
\[
\Theta_n(\rho)=\frac{1}{\sqrt{N}}\int_{-\infty}^\infty\ldots\int_{-\infty}^\infty
\exp\left(\beta Jd \rho^2N+\beta J{\sum_{k\in V_n}}'\lambda_k
y_k^2+\beta\rho{\sum_{j\in V_n}}h_j+\beta{\sum_{k\in
V_n}}'\varphi_k y_k\right)\times
\]
\[
\delta\left({\sum_{k\in V_n}}'y_k^2-N(1-\rho^2)\right){\prod_{k\in
V_n}}'dy_k,
\]
where primes indicate that summations/products do not include
$k=(1,1,\ldots,1)$.

The integral representation for the delta function
\[
\delta\left({\sum_{k\in
V_n}}'y_k^2-N(1-\rho^2)\right)=\frac{1}{2\pi
i}\int_{-i\infty}^{+i\infty}\!ds\,
\exp\left[s\left(N(1-\rho^2)-{\sum_{k\in
V_n}}'y_k^2\right)\right],
\]
allows one to perform integration over the variables $y_k$.
However, we can switch the order of integration over the variables
$y_k$ and $s$ only after a shift of the integration contour for
$s$ to the right. The shift must assure that the real part of
the quadratic form involving the variables $y_k$, is negatively
defined. On switching the integration order, integrating over
$y_k$, and introducing a new integration variable $z$ via
$s=\beta J z$ one obtains
\begin{equation}
\Theta_n(\rho)=\frac{\beta J}{2\pi i\sqrt{N}}\exp\left(\beta Jd
\rho^2N+\beta\rho{\sum_{j\in V_n}}h_j\right)\left(\frac{\pi}{\beta
J}\right)^{(N-1)/2}\int_{-i\infty+c}^{+i\infty+c}\!dz\,
\exp\left[N\beta\Phi_n(z)\right], \label{pf1}
\end{equation}
where
\[
\Phi_n(z)=J z(1-\rho^2)-\frac{1}{2\beta N}{\sum_{k\in
V_n}}'\ln(z-\lambda_k)+\frac{1}{4J N}{\sum_{k\in
V_n}}'\frac{\varphi_k^2}{z-\lambda_k},
\]
and $c>d$ is the shift of the integration contour mentioned above.

Depending on the value of $\beta$, the large-$n$ asymptotics of
the integral (\ref{pf1}) can be found either by the saddle-point
method, or by direct integration after an appropriate change of
variables. The saddle point of the integrand is a solution of the
equation
\begin{equation}
\Phi_n'(z)=J(1-\rho^2) -\frac{1}{2\beta N}{\sum_{k\in
V_n}}'\frac{1}{z-\lambda_k}-\frac{1}{4J N}{\sum_{k\in
V_n}}'\frac{\varphi_k^2}{(z-\lambda_k)^2}=0. \label{sp1}
\end{equation}

If the field $\{h_{j},j\in Z^d\}$ is of the type (\ref{ft}), then
\[
\frac{1}{N}{\sum_{k\in
V_n}}\frac{\varphi_k^2}{(z-\lambda_k)^2}\leq
\frac{n^{-2\gamma}b^2}{(z-d)^2}.
\]
Hence, for any $z>d$, as $n\to\infty$, the sequence of the
derivatives $\Phi_n'(z)$ converges to
\[
\Phi'(z)= J(1-\rho^2)-\frac{1}{2\beta}W_d(z),
\]
where
\[
W_d(z)\equiv\int_{-\pi}^{\pi}\ldots \int_{-\pi}^{\pi}
\frac{1}{z-\sum_{\nu=1}^d \cos \omega_\nu}\prod_{l=1}^d
\frac{d\omega_l}{2\pi}< \infty \label{wf}
\]
is the Watson function.

The function $\Phi'(z)$ increases monotonically with $z$ on
$[d,\infty)$, and the location of its zeroes depends on the
dimension $d$ of the lattice. Namely, if $d=1,2$, then the
function $\Phi'(z)$ has exactly one zero on the interval
$[d,\infty)$ at a point $z^*(\rho)>d$, for any $\beta>0$. If
$d\geq 3$, then there exists a critical value
\begin{equation}
\beta_c(\rho) =\frac{1}{2J(1-\rho^2)}W_d(d)
\label{ct}
\end{equation}
of the parameter $\beta$. If $\beta < \beta_c(\rho)$, then the
function $\Phi'(z)$ still has exactly one zero on the interval
$[d,\infty)$ at a point $z^*>d$. While if $\beta > \beta_c(\rho)$,
then the function $\Phi'(z)$ is strictly positive on the interval
$[d,\infty)$.

In this section our aim is to investigate the natural state of the
spherical lattice gas, which, as we shall see shortly, happens to
be a mixed phase. Therefore we do not utilize any symmetry-breaking
perturbations and set $h_j=0$ for all $j\in Z^d$ untill
Section 6.

Application of the saddle-point method to the integral
(\ref{pf1}) is fairly straightforward when there exists a saddle
point $z^*(\rho)$ greater than $d$. In this case
\[
\Phi_n(z^*(\rho))=\underbrace{J(1-\rho^2)z^*(\rho)-\frac{1}{2\beta}L_d(z^*(\rho))}_{\equiv
\Phi(z^*(\rho))}+O(e^{-n\delta}),
\]
and the saddle-point method yields
\[
-f_n\equiv\frac{1}{\beta n^d}\ln\Theta_n(\rho)=
\frac{1}{2\beta}\ln\frac{\pi}{\beta
J}+Jd\rho^2+\Phi(z^*(\rho))+O\left(n^{-d}\ln n\right),
\]
as $n\to\infty$, where
\[
L_d(z)=\int_{-\pi}^{\pi}\!\ldots \int_{-\pi}^{\pi}
\ln\left(z-\sum_{\nu=1}^d \cos \omega_\nu\right)\prod_{l=1}^d
\frac{d\omega_l}{2\pi}.
\]

When $\beta\geq\beta_c$, the function $\Phi_n(z)$ still attains
its minimum on the interval $(\lambda_{\rm \,s},\infty)$ at a
point $z_n^*>\lambda_{\rm\, s}$, where $\lambda_{\rm\,
s}=d-1+\cos\frac{2\pi}{n}$ is the second-largest eigenvalue of the
interaction matrix $\widehat T$. However, the sequence of saddle
points $z_n^*$ approaches the pole of the integrand at
$z=\lambda_{\rm\, s}$, and the application of the saddle-point
method becomes a bit more tricky.

To find the asymptotics of the free energy we have to introduce a
new integration variable $\zeta$ via $z=\lambda_{\rm\,s}+\zeta
N^{-1}$. The large-$n$ asymptotics of the sum
\[
L_{d}^{(N)}\left(\lambda_{\rm\,s}+\zeta
N^{-1}\right)\equiv\frac{1}{N}{\sum_{k\in V_n}}'
\ln\left(\lambda_{\rm\,s}+\zeta N^{-1}-\lambda_k\right),
\]
is given by
\[
L_{d}^{(N)}\left(\lambda_{\rm\,s}+\zeta N^{-1}\right)=
\frac{2d\ln(\zeta/ N)}{N}+\frac{1}{N}{\sum_{k\in V_n}}''
\ln\left(\lambda_{\rm\,s}-\lambda_k\right)
-\frac{\zeta}{N^2}{\sum_{k\in V_n}}''
\frac{1}{\lambda_{\rm\,s}-\lambda_k}+o(N^{-1}),
\]
where the double prime indicates that the summation does not
include both the largest and the ($2d$-times degenerate)
second-largest eigenvalues.

Hence
\[
\Theta_n(\rho)\sim\exp\left[N\beta
J(d\rho^2+\lambda_{\rm\,s}(1-\rho^2))-\frac{1}{2} {\sum_{k\in
V_n}}'' \ln\left(\lambda_{\rm\,s}-\lambda_k\right)\right]\times
\]
\begin{equation}
\beta
JN^{d-3/2}\left(\frac{\pi}{\beta J}\right)^{(N-1)/2}\frac{1}{2\pi i}\int_{C}
\frac{e^{J[(1-\rho^2)\beta-\beta_c]\zeta}}{\zeta^d}\,d\zeta,
\label{pfsg}
\end{equation}
where the integration contour $C$ runs below the negative
semi-axis $(-\infty,0]-i\,0$, encircles $0$ counterclockwise, and returns
to $-\infty$ above the negative semi-axis $(-\infty,0]+i\,0$.

We see that the remaining integration does not contain the large
parameter $N$ any longer, and calculating the residue of the
integrand at $\zeta=0$ we obtain
\[
\frac{1}{2\pi i}\int_{C}
\frac{e^{J[(1-\rho^2)\beta-\beta_c]\zeta}}{\zeta^d}\,d\zeta=
\frac{J^{d-1}}{(d-1)!}\left[(1-\rho^2)\beta-\beta_c\right]^{d-1}.
\]

We can use the same method to find the joint characteristic
function $\chi(t,s)$ of random variables $x_j$ and $x_l$. First,
we express $\chi(t,s)\equiv\langle \exp(itx_j+isx_l)\rangle_n$ as
the following integral
\[
\chi(t,s)=\frac{e^{it\rho+is\rho}}{\Theta_n(\rho)\sqrt{N}}
\exp\left(\beta Jd\rho^2N+\beta\rho\sum_{j\in
V_n}h_j\right)\frac{\beta J}{2\pi i}\left(\frac{\pi}{\beta
J}\right)^{(N-1)/2}\times
\]
\begin{equation}
\int_{-i\infty+c}^{+i\infty+c}\!
dz\,\exp\left[N\beta\Phi_n(z)-{\sum_{k\in V_n}}'
\frac{\left(tw_j^{(k)}+sw_l^{(k)}\right)^2}{4\beta
J(z-\lambda_k)}+i{\sum_{k\in V_n}}'
\frac{\left(tw_j^{(k)}+sw_l^{(k)}\right)\varphi_k}{2J(z-\lambda_k)}
\right]. \label{chi1}
\end{equation}
Next, introduce a new integration variable $\zeta$ via
$z=\lambda_{\rm\,s}+\zeta N^{-1}$, and use the asymptotic
expansion for the partition function $\Theta_n(\rho)$ to obtain
\[
\chi(t,s)\sim\exp\left[i(t+s)\rho-\frac{t^2+s^2}{4\beta
JN}{\sum_{k\in V_n}}'' \frac{1}{\lambda_{\rm\,
s}-\lambda_k}-\frac{ts}{2\beta J}{\sum_{k\in V_n}}''
\frac{w_j^{(k)}w_l^{(k)}}{\lambda_{\rm\,
s}-\lambda_k}\right]\times
\]
\begin{equation}
\frac{(d-1)!}{2\pi i}\int_{C}
\exp\left[\zeta-\frac{1-\rho^2}{2\,\zeta}
\left(1-\frac{\beta_c}{\beta}\right) \left((t^2+s^2)d+
2\,ts\sum_{\nu=1}^d\cos\frac{2\pi(j_\nu-k_\nu)}{n}\right)\right]
\, \frac{d\zeta}{\zeta^d}. \label{chf1}
\end{equation}
The remaining integral can be expressed in terms of the Bessel
function
\[
J_m(x)=\frac{1}{2\pi i}\int_{C}
\exp\left[\frac{x}{2}\left(\zeta-
\frac{1}{\zeta}\right)\right] \, \frac{d\zeta}{\zeta^{m+1}},
\]
however, calculation of the moments can be done easily by
differentiation under the integral sign in Eq.\ (\ref{chf1}).

Setting $s=0$ and passing to the limit $N\to\infty$ we see that
the random variables $x_j$ of the spherical model on the infinite
lattice have the common characteristic function
\begin{equation}
\chi(t)=\exp\left[it\rho-\frac{t^2}{4\beta J}W_d(d)\right]
\frac{(d-1)!}{2\pi i}\int_{C}
\exp\left[\zeta-\frac{1-\rho^2}{2\,\zeta}
\left(1-\frac{\beta_c}{\beta}\right) t^2d\right] \,
\frac{d\zeta}{\zeta^d}. \label{chfi}
\end{equation}
Differentiation over $t$ shows that the expected values are given
by $\langle x_j\rangle=\rho$, and the variances are given by
\[
\mbox{Var}(x_j)=\frac{W_d(d)}{2\beta
J}+\frac{1-\rho^2}{2}\left(1-\frac{\beta_c}{\beta}\right).
\]
More importantly, differentiation of the joint characteristic
function shows that
\[
\mbox{Cov}(x_j,x_l)=
(1-\rho^2)\left(1-\frac{\beta_c}{\beta}\right)
\frac{1}{d}\sum_{\nu=1}^d\cos\frac{2\pi(j_\nu-l_\nu)}{n}.
\]
Note that $\mbox{Cov}(x_j,x_l)\not\to0$, as
$\mbox{dist}(j,l)\to\infty$, for any $\beta>\beta_c$. Therefore,
the lattice gas is in a mixed (or degenerate) state, and in order
to calculate the values of observable quantities we have to single out
pure phases.

In the case of the stretched lattice (\ref{stretch}) the
second-largest eigenvalue of the interaction matrix is only twice degenerate,
and we obtain
\[
\chi(t,s)\sim\exp\left[i(t+s)\rho-\frac{t^2+s^2}{4\beta
JN}{\sum_{k\in \Upsilon_n}}'' \frac{1}{\lambda_{\rm\,
s}-\lambda_k}-\frac{ts}{2\beta J}{\sum_{k\in \Upsilon_n}}''
\frac{w_j^{(k)}w_l^{(k)}}{\lambda_{\rm\,
s}-\lambda_k}\right]\times
\]
\begin{equation}
J_0\left[\sqrt{2(1-\rho^2) \left(1-\frac{\beta_c}{\beta}\right)
\left(t^2+s^2+
2\,ts\cos\frac{2\pi(j_1-l_1)}{n(1+\delta)}\right)}\right],
\label{chf1s}
\end{equation}
where $J_0(z)$ is the Bessel function of order zero. Hence, for
$1\ll |j_1-l_1|\ll n$ we have
\[
\chi(t,s)\sim\mbox{\boldmath $E$}\exp\left[i(t+s)\sqrt{2(1-\rho^2)
\left(1-\frac{\beta_c}{\beta}\right)}\sin 2\pi\mbox{\boldmath
$U$}+\right.
\]
\[
\left.+it\,{\cal N}_j\left(\rho,\frac{W_d(d)}{2\beta
J}\right)+is\,{\cal N}_l\left(\rho,\frac{W_d(d)}{2\beta
J}\right)\right],
\]
where the random variable \mbox{\boldmath $U$} is uniformly
distributed on the interval $[0,1]$, and it is independent of the normal
random variables ${\cal N}_j$ and ${\cal N}_l$.

The last expression admits the following probabilistic
interpretation. Let
\[
j/n\equiv\left(\frac{j_1}{n},\frac{j_2}{n},\ldots,\frac{j_d}{n}\right)
\to\gamma\equiv(\gamma_1,\gamma_2,\ldots,\gamma_d),\ \mbox{as}\ n\to\infty,
\]
then in the continuous limit the field of random variables $x_j$ converges
to the random function
\[
x(\gamma)=\sqrt{2(1-\rho^2)
\left(1-\frac{\beta_c}{\beta}\right)}\sin 2\pi\mbox{\boldmath
$U$}_{\gamma}+{\cal N}_\gamma\left(\rho,\frac{W_d(d)}{2\beta
J}\right),
\]
where ${\cal N}_\gamma\left(a,b\right)$ is a continuous function of
independent normal random variables defined on the rectangle
\[
0\leq\gamma_1\leq 1+\delta,\ 0\leq\gamma_2\leq 1,\ldots,0\leq\gamma_d\leq 1.
\]
The random variables $U_\gamma$ are uniformly distributed on the interval
$[0,1]$, any pair $U_\gamma$, $U_\alpha$ becomes perfectly correlated as
$\gamma_1\to\alpha_1$. While if $\gamma_1\neq\alpha_1$, the random variables
$U_\gamma$, $U_\alpha$ are neither independent, nor perfectly correlated.

\section{Droplet shape at zero temperature.}

As it often happens, the situation at zero temperature is simpler
than in the case $\beta^{-1}>0$. If $\beta^{-1}=0$, then the pure
states of the spherical gas are configurations $\mbox{\boldmath
$x$}_N=\{x_j,j\in V_n\}$ which minimize the energy $H_n$ subject
to the spherical and the density constraints. The corresponding
minimization problem is solved using the Lagrange function
\[
{\cal L}(\{x_j,j\in V_n\},a,b)=-J\sum_{j,k\in V_n}T_{jk}x_j
x_k+a\left(\sum_{j\in V_n}x_j^2-N\right)+b\left(\sum_{j\in
V_n}x_j-\rho N\right),
\]
where $a$ and $b$ are Lagrange multipliers. Introducing the
variables $\{y_j,j\in V_n\}$, see Eq.\ (\ref{nc}), we get rid of
the density constraint, and write down the Lagrange function in
the following diagonal form
\[
{\cal L}(\{y_j,j\in V_n\},a)=-JN\rho^2\lambda_{(1,1,\ldots,1)}-
J{\sum_{j\in V_n}}'\lambda_j y_j^2 +a\left({\sum_{j\in
V_n}}'y_j^2-N(1-\rho^2)\right),
\]
where the primes indicate that the summations do not involve
$j=(1,1,\ldots,1)$.

Differentiating over $\{y_j,j\in V_n\setminus (1,1,\ldots,1)\}$ we
obtain the following system of equations for stationary points
\[
2J\lambda_j y_j=2ay_j,\quad j\in V_n\setminus(1,1,\ldots,1).
\]
If $a\neq J\lambda_k$, then the only solution of the system,
$y_j=0$ for $j\in V_n\setminus(1,1,\ldots,1)$, violates the
spherical constraint. Therefore to solve the constraint
minimization problem we have to set the value of the Lagrange
multiplier to $a=J\lambda_k$, for some $k\in V_n\setminus
(1,1,\ldots,1)$. Using the spherical constraint we obtain for such
a value of $a$, that
\[
\sum_{j:J\lambda_j=a}y_j^2=N(1-\rho^2),\quad\mbox{and} \quad
y_j=0, \quad\mbox{if}\quad J\lambda_j\neq a.
\]

The global minimum of $-\sum_{j\in V_n}'\lambda_j y_j^2$ is
obtained when $a=J\lambda_{\,\rm s}$, where $\lambda_{\,\rm s}$ is
the $2d$ times degenerate second largest eigenvalue of the
interaction matrix. Therefore the configuration $\{y_j^*,j\in
V_n\}$ minimizing $-{\sum_{j\in V_n}}\lambda_j y_j^2$ subject to the
spherical and density constraints has the following components.
The component $y_{(1,1,\ldots,1)}^*=\rho\sqrt{N}$, due to the
density constraint. The $2d$ components $y_j^*$, with $j$ such
that $\lambda_j=\lambda_{\,\rm s}$, are the coordinates an arbitrary
point on a $2d$-dimensional sphere with the center at the origin and the
radius $\sqrt{N(1-\rho^2)}$. All the remaining $N-2d-1$ components
of the configuration $\{y_j^*,j\in V_n\}$ are zeroes.

Going back to the variables $\{x_j,j\in V_n\}$ we obtain
\[
x_j=w_j^{(1,1,\ldots,1)}\rho\sqrt{N}+\sum_{k:\lambda_k=\lambda_{\,\rm
s}}w_j^{(k)}y_k^*=\rho+\sum_{k:\lambda_k=\lambda_{\,\rm
s}}w_j^{(k)}y_k^*.
\]
The $2d$ orthonormal eigenvectors corresponding to the eigenvalue
$\lambda_{\,\rm s}$ are given by
\[
\mbox{\boldmath
$w$}^{\pm}_\nu=\left\{\sqrt{\frac{2}{N}}\cos\left[\frac{2\pi
(j_\nu-1)}{n}\pm\frac{\pi}{4}\right]\right\}_{j\in V_n},\quad
\nu=1,2,\ldots,d.
\]
Let us denote $y^\pm_\nu$ the component $y^*_k$ such that
$\mbox{\boldmath $w$}^{(k)}=\mbox{\boldmath $w$}^{\pm}_\nu$. Then
the components of pure zero-temperature states $\mbox{\boldmath
$x$}_N$ are given by
\begin{eqnarray}
x_j&=&\rho+\sqrt{\frac{2}{N}}\sum_{\nu=1}^d\cos\left[\frac{2\pi
(j_\nu-1)}{n}-\frac{\pi}{4}\right]y^-_\nu+\cos\left[\frac{2\pi
(j_\nu-1)}{n}+\frac{\pi}{4}\right]y^+_\nu \nonumber\\
&=&\rho+\sqrt{\frac{2}{N}}\sum_{\nu=1}^d\sqrt{(y^-_\nu)^2+(y^+_\nu)^2}
\,\cos\left[\frac{2\pi(j_\nu-1)}{n}-\alpha_\nu\right],
\label{evst0}
\end{eqnarray}
where the phase shifts $\alpha_\nu$ are such that
\[
\cos\left(\alpha_\nu-\frac{\pi}{4}\right)=
\frac{y_\nu^-}{\sqrt{(y^-_\nu)^2+(y^+_\nu)^2}},\quad
\mbox{and}\quad
\sin\left(\alpha_\nu-\frac{\pi}{4}\right)=
-\frac{y_\nu^-}{\sqrt{(y^-_\nu)^2+(y^+_\nu)^2}}.
\]

The phase shifts $\{\alpha_\nu\}_{\nu=1}^d$ merely set the
``center" of the pure state $\mbox{\boldmath $x$}_N$, which we are
free to choose at will because of the translation invariance of
the Hamiltonian of the spherical lattice gas. The multipliers
$\sqrt{(y^-_\nu)^2+(y^+_\nu)^2}$ play a much more important role,
they determine asymmetry of the droplet across dimensions. If
$(y^-_k)^2+(y^+_k)^2=N(1-\rho^2)$, and $y^\pm_\nu=0$, for $\nu\neq
k$, then the states are maximally asymmetric. Namely, they have the
cosine shape in the dimension $\nu=k$, and they are translation
invariant across other dimensions, see Fig.\ 1,
\[
x_j=\rho+\sqrt{2(1-\rho^2)}
\cos\left[\frac{2\pi(j_k-1)}{n}-\alpha_k\right].
\]

The opposite extreme is the symmetric case
$(y^-_\nu)^2+(y^+_\nu)^2=N(1-\rho^2)/d$, for $\nu=1,2,\ldots,d$,
when the droplet has a rounded-square shape, see Fig.\ 1,
\begin{equation}
x_j=\rho+\sqrt{\frac{2(1-\rho^2)}{d}}
\sum_{\nu=1}^d\cos\left[\frac{2\pi(j_\nu-1)}{n}-\alpha_\nu\right].
\label{tocf}
\end{equation}
\begin{figure}[htb]
\begin{center}
\epsfig{file=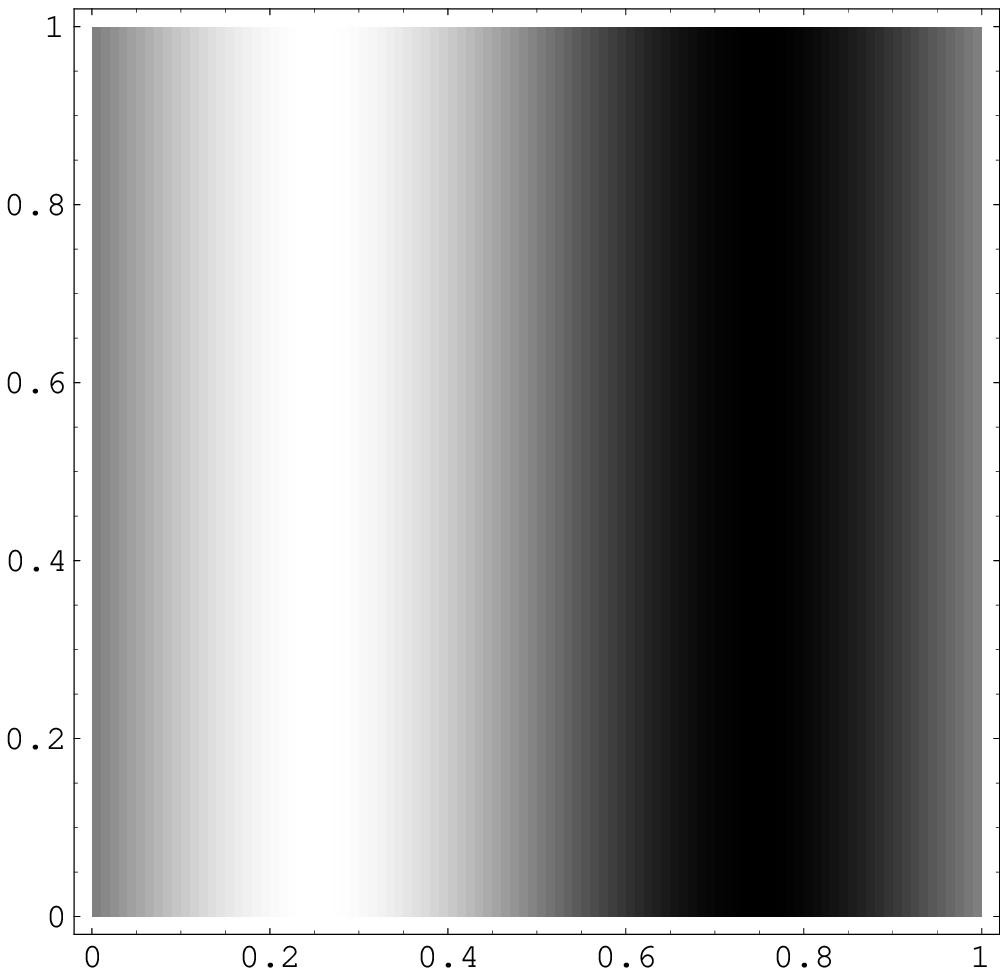, height=6cm,width=6cm}
\epsfig{file=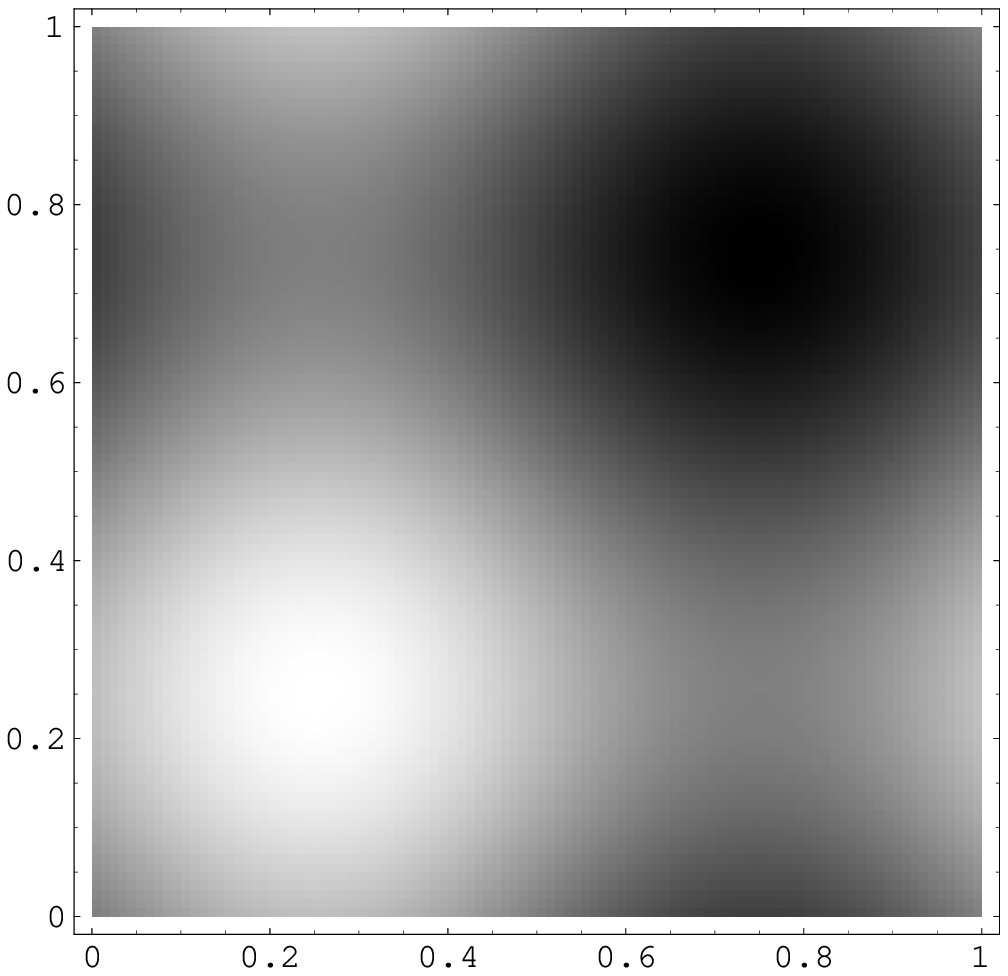, height=6cm,width=6cm}
\end{center}
\hskip -2cm \vspace*{-1.2cm} \caption{The shape of the maximally
asymmetric and symmetric droplets. \label{Fig. 1}}
\end{figure}

\noindent In intermediate situations, the droplet has a kind of
elliptic shape, see Fig.\ 2.

\begin{figure}[htb]
\begin{center}
\epsfig{file=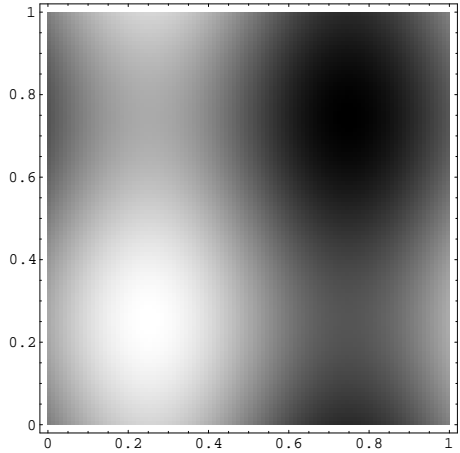, height=6cm,width=6cm}
\end{center}
\hskip -2cm \vspace*{-1.2cm} \caption{A droplet of an intermediate
shape. \label{Fig. 2}}
\end{figure}

It turns out that the above zero-temperature pure (macro) states
are stable with respect to heating up of the system to a
temperature not exceeding the critical value $\beta_c^{-1}$. Pure
(macro) states for $\beta^{-1}>0$ are in one-to-one
correspondence with zero-temperature pure states. In the next section
we single out pure states for $\beta^{-1}>0$ by quasi-averaging.

\section{Quasi-averages.}

Pure (micro) states of thermodynamic systems can be obtained with
a help of quasi-averaging \cite{b60}. The basic recipe of
quasi-averaging looks like this: switch on an external field (of
magnitude $\varepsilon$) removing the degeneracy of equilibrium
state, pass to the thermodynamic limit, switch off the field by
sending its magnitude $\varepsilon$ to zero over positive values
($\varepsilon\downarrow0$). It turns out that in the case of the
spherical lattice gas this procedure often deforms macro states
(droplet shape). Therefore, we have
to use a more potent version of quasi-averaging \cite{az85,bzt86},
where the field magnitude is a decreasing function
of the volume $N$. That is, the symmetry-breaking external field
is switched off together with the thermodynamic limit.

Motivated by the results of the previous section, we begin with
the Hamiltonian (\ref{ham}) where the external field
\begin{equation}
h_j=\frac{\varepsilon\sqrt{2}}{n^{\delta}}
\cos\left[\frac{2\pi(j_1-1)}{n}-\alpha\right],\quad
j\equiv(j_1,j_2,\ldots,j_d)\in V_n, \label{qah}
\end{equation}
is proportional to an eigenvector corresponding to the
second-largest eigenvalue, $\lambda_{\,\rm s}$, of the interaction
matrix. The dimensionality of the eigenspace corresponding to
$\lambda_{\,\rm s}$ is $2d$, and we are free to choose the value of
the parameter $\alpha\in[0,2\pi)$, and instead of $j_1$ we can use
$j_\nu$, with $\nu=2,3,\ldots,d$. Different choices of $\alpha$
and $\nu$ yield different pure macro states.

For the characteristic function of random variables $x_j$ and
$x_l$ one obtains an integral similar to Eq.\ (\ref{chi1}),
\[
\chi(t,s)=\frac{\exp(\beta
Jd\rho^2N)e^{it\rho+is\rho}}{\Theta_n(\rho)\sqrt{N}}\frac{\beta J}{2\pi
i}\left(\frac{\pi}{\beta J}\right)^{(N-1)/2}\times
\]
\begin{equation}
\int_{-i\infty+c}^{+i\infty+c}\!
dz\,\exp\left[N\beta\Phi_n(z,\varepsilon)\right]\exp\left[-{\sum_{k\in
V_n}}' \frac{\left(tw_j^{(k)}+sw_l^{(k)}\right)^2}{4\beta
J(z-\lambda_k)}\right]. \label{chi2}
\end{equation}
where
\[
N\beta\Phi_n(z,\varepsilon)=N\beta\Phi_n(z)
+\frac{i\,\varepsilon}{2Jn^{\delta-d/2}}\frac{tw_j^{(2,1,\ldots,1)}+
sw_l^{(2,1,\ldots,1)}}{z-\lambda_{\rm\,s}}+
\underbrace{\frac{\beta\varepsilon^2}{4Jn^{2\delta-d}(z-\lambda_{\rm\,s})}}_{\equiv
\psi_n(z)}.
\]
The innocent-looking term $\psi_n(z)$ makes a huge difference for
the evaluation and behaviour of the integral (\ref{chi2}) when
$\beta>\beta_c$. Indeed, in the scale $z=\lambda_{\rm\,s}+\zeta
n^{-\delta}$ the magnitudes of variation of $N\beta\Phi_n(z)$ and
$\psi_n(z)$ with $\zeta$ become comparable. Moreover, if
$\delta<d$, then $\psi_n(z)$ prevents the saddle point $z_n^*$
from approaching the pole at $\lambda_{\rm\,s}$ any closer than
the distance $O(n^{-\delta})$.

Introducing  a new integration variable $\zeta$ via
$z=\lambda_{\rm\,s}+\zeta n^{-\delta}$ one obtains
\[
N\beta\Phi_n(\lambda_{\rm\,s}+\zeta
n^{-\delta},\varepsilon)=N\beta\Phi_n(\lambda_{\rm\,s}+\zeta
n^{-\delta})+
\]
\[
i\frac{\varepsilon\sqrt{2N}}{2J\zeta}\left\{t\cos
\left[\frac{2\pi(j_1-1)}{n}-\alpha\right]+ s\cos
\left[\frac{2\pi(l_1-1)}{n}-\alpha\right]\right\}+
\frac{n^{d-\delta}\beta\varepsilon^2}{4J\zeta}.
\]
Gathering the terms of the order $n^{d-\delta}$ and
differentiating over $\zeta$ we obtain the following equation for
the saddle-point $\zeta^*$:
\[
(1-\rho^2)J(\beta-\beta_c)-\frac{\beta\varepsilon^2}{4J\zeta^2}=0.
\]
Hence the saddle point of $\Phi_n(z,\varepsilon)$ is located at
\[
z_n^*=\lambda_{\rm\,s}+\frac{|\varepsilon|}{2Jn^\delta}
\frac{1}{\sqrt{(1-\rho^2)(1-\beta_c/\beta)}}+o(n^{-\delta}).
\]

Application of the saddle-point method to the integral
(\ref{chi2}) and to the analogous integral for the partition
function $\Theta_n(\rho)$ shows that the large-$n$ asymptotics of
$\chi(t,s)$ coincides with the characteristic function of a
bivariate normal distribution ${\cal B}(\mu_1,\mu_2;v_1,v_2,c)$.
That is, for large $n$ the joint distribution of random variables
$x_j$ and $x_l$ is asymptotically normal with mean values $\mu_1$ and
$\mu_2$, variances $v_1^2$ and $v_2^2$, respectively, and
covariance $c$. Under the assumption that the first components of
the nodes $j$ and $l$ scale with $n$ as $j_1\sim \gamma n$ and
$l_1\sim\lambda n$, the expected values of $x_j$ and $x_l$ are
given by
\begin{eqnarray}
&&\mu_1=\rho+\sqrt{2(1-\rho^2)(1-\beta_c/\beta)}\cos\left(2\pi
\gamma-\alpha\right),\label{shape1}\\
&&\mu_2=\rho+\sqrt{2(1-\rho^2)(1-\beta_c/\beta)}\cos\left(2\pi
\lambda-\alpha\right).\label{shape2}
\end{eqnarray}
Note that, provided $\delta\in(0,d)$, the expected values (droplet
shape) do not depend on the magnitude of the external field
$\varepsilon n^{-\delta}$. Therefore it is reasonable to conclude
that the symmetry-breaking field (\ref{qah}) selects one of the possible
droplet shapes, but it does not deform the droplet.

The variances $v_1^2$, $v_2^2$ and covariance $c(j,l)$ are given
by the usual formulae for the pure phases of the spherical model
below the critical temperature
\[
v_1^2=v_2^2=\frac{W_d(d)}{2\beta J},
\]
\begin{equation}
c(j,l)=\frac{1}{2\beta J}\int_{-\pi}^{\pi}\ldots\int_{-\pi}^{\pi}
\frac{\exp\left[i\sum_{\nu=1}^d (j_\nu-l_\nu)\omega_\nu\right]}
{d-\sum_{\nu=1}^d \cos\omega_\nu}\prod_{\nu=1}^d
\frac{d\omega_\nu}{2\pi}\sim\frac{\Gamma(d/2-1)}{4\beta
J\pi^{d/2}r_{j,l}^{d-2}},
\label{ucf}
\end{equation}
where $r_{j,l}$ is the Euclidean distance between $j$ and $l$.

The equations (\ref{shape1}) and (\ref{shape2}) describe the shape of
a localised droplet obtained with a help of the (generalized, see
\cite{az85,bzt86}) method of quasi-averages. To double check that the
droplet shape (\ref{shape1}) is not deformed by the external
field we would like to examine the sensitivity of the droplet shape to the type
of external field used for quasi-averaging. For this purpose we now repeat the
above calculations for a technically more demanding case
\begin{equation}
h_j=\left\{
\begin{array}{cl}
\varepsilon n^{-\delta},&\mbox{ if }j_1=1, \\
0,&\mbox{ otherwise,}
\end{array}
\right. \qquad j\in V_n; \label{sbf}
\end{equation}
where $\varepsilon>0$, and $\delta\geq0$. First, we have to find the range
of values for $\delta$, such that the field (\ref{sbf}) is strong enough
to fix the location of the droplet of condensed ``spherical matter",
but, at the same time, it is weak enough not to deform the droplet shape.

The partition function
$\Theta_n(\rho)$ is still given by Eq.\ (\ref{pf1}), but
the coefficients $\varphi_k$, $k\in V_n$ are now given by
\[
\varphi_k=\varepsilon
n^{d/2-1-\delta}\prod_{\nu=2}^d\delta(k_\nu,1).
\]
In order to find the location of the saddle-point of the integrand
in Eq.\ (\ref{pf1}) we have to first investigate the behaviour of the
sum
\[
\Sigma(z)\equiv\frac{1}{4JN}{\sum_{k\in
V_n}}'\frac{\varphi_k^2}{z-\lambda_k}=
\frac{\varepsilon^2}{4Jn^{2(1+\delta)}}\sum_{k_1=2}^n
\frac{1}{z-d+1-\cos\left[2\pi(k_1-1)/n\right]}
\]
in the vicinity of the point $z=d$. Fortunately, this sum can be calculated
exactly, see \cite{p94},
\[
\Sigma(z)=
\frac{\varepsilon^2}{4Jn^{2(1+\delta)}}\left[\frac{2n}{x_2(z)-x_1(z)}
\frac{x_2^n(z)+1}{x_2^n(z)-1}-\frac{1}{z-d}\right].
\]
Hence, for any fixed $z>d$, the (derivative of the) sum
$\Sigma(z)=O(n^{-1-2\delta})$ produces only a vanishing
contribution to the saddle-point equation (\ref{sp1}).

The contribution of the sum $\Sigma(z)$ becomes significant below
the critical temperature, where we have to introduce a new
integration variable $\zeta$ via $z=\lambda_{\,\rm s}+\zeta
n^{-\gamma}$ before application of the saddle-point method.
In order to find the right rescaling for the integration variable
$z$ (which, as we shall see shortly, depends on the
value of $\delta$ in Eq.\ (\ref{sbf})), we have to analyse the behavior
of $\Sigma(z)$ for different values of $\gamma$.

If
$0<\gamma<2$, then
\[
\Sigma(\lambda_{\,\rm s}+\zeta n^{-\gamma})\sim
\frac{\varepsilon^2}{4J\sqrt{2\zeta}}
\,n^{-1-2\delta+\gamma/2},\quad\mbox{as $n\to\infty$.}
\]

If $\gamma=2$, then
\[
\Sigma(\lambda_{\,\rm s}+\zeta n^{-\gamma})\sim \frac{
\varepsilon^2n^{-2\delta}}{8J}
\left(\frac{\coth\sqrt{\zeta/2-\pi^2}}{\sqrt{\zeta/2-\pi^2}}-
\frac{1}{\zeta/2-\pi^2}\right), \quad\mbox{as $n\to\infty$.}
\]
Note that the r.h.s.\ of the last equation does not actually have a
singularity at $\zeta=2\pi^2$. Therefore to find
$\Sigma(\lambda_{\,\rm s}+\zeta n^{-\gamma})$ for
$\zeta<2\pi^2$ one can use
the analytic continuation
\[
\Sigma(\lambda_{\,\rm s}+\zeta n^{-\gamma})\sim -\frac{
\varepsilon^2n^{-2\delta}}{8J}
\left(\frac{\cot\sqrt{\pi^2-\zeta/2}}{\sqrt{\pi^2-\zeta/2}}+
\frac{1}{\zeta/2-\pi^2}\right), \quad\mbox{as $n\to\infty$.}
\]

Finally, if $\gamma>2$, then the main asymptotics of
$\Sigma(\lambda_{\,\rm s}+\zeta n^{-\gamma})$ comes entirely from
the first and the last terms of this sum,
\[
\Sigma(\lambda_{\,\rm s}+\zeta n^{-\gamma})\sim \frac{
\varepsilon^2n^{\gamma-2(1+\delta)}}{2J\zeta},\quad\mbox{as
$n\to\infty$.}
\]

Under the same rescaling $z=\lambda_{\,\rm s}+\zeta n^{-\gamma}$
the first two terms of $\Phi_n(z)$ become
\[
J(\lambda_{\,\rm s}+\zeta n^{-\gamma})(1-\rho^2)-\frac{1}{2\beta
N}{\sum_{k\in V_n}}'\ln(\lambda_{\,\rm s}+\zeta
n^{-\gamma}-\lambda_k)\sim
\]
\[
J\lambda_{\,\rm s}(1-\rho^2)-\frac{L_d(d)}{2\beta}+
n^{-\gamma}\zeta
\left[J(1-\rho^2)-\frac{W_d(d)}{2\beta}\right]+n^{-2}\frac{\pi^2W_d(d)}{\beta}.
\]
Hence, if $0<\delta<1$, then the contributions of all $\zeta$-dependent
terms in $\Phi_n(\lambda_{\,\rm s}+\zeta n^{-\gamma})$
are of the same order when $\gamma=2(1+2\delta)/3$, and the
saddle-point equation for $\zeta$ is given by
\[
J(1-\rho^2)(\beta-\beta_c)=\frac{\beta}{4J}\frac{
\varepsilon^2}{2\sqrt{2}\zeta^{3/2}}.
\]
The positive solution of this equation is given by
\begin{equation}
\zeta^*=\frac{1}{2}\left(\frac{\varepsilon}{2J}\right)^{4/3}
\frac{1}{\left[(1-\beta_c/\beta)(1-\rho^2)\right]^{2/3}}.
\label{sp2}
\end{equation}

If $\delta=1$, then we have to set $\gamma=2$, which yields the
following saddle-point equation for $\zeta$
\begin{equation}
J(1-\rho^2)(\beta-\beta_c)=-\frac{\beta
\varepsilon^2}{8J}\frac{d}{d\zeta}\left(
\frac{\coth\sqrt{\zeta/2-\pi^2}}{\sqrt{\zeta/2-\pi^2}}
-\frac{1}{\zeta/2-\pi^2}\right). \label{sp3}
\end{equation}

Finally, if $1<\delta<d-1$, then we have to set $\gamma=1+\delta$,
and the saddle-point equation for $\zeta$ is given by
\[
J(1-\rho^2)(\beta-\beta_c)=\frac{\beta \varepsilon^2}{2J\zeta^2}.
\]
The positive solution of the above equation is given by
\[
\zeta^*=\frac{\varepsilon}{J\sqrt{2(1-\beta_c/\beta)(1-\rho^2)}}.
\]

Now that we know the behavior of the saddle point $z_n^*$
of the integrand in Eq.\ (\ref{pf1}) we can find the thermodynamic
limits of various macro- and microscopic quantities. But before we
are able to apply the
saddle-point method and find the characteristic function of an arbitrary
pair of random variables $x_j$ and $x_l$ we have to calculate the sum
\[
\widetilde\Sigma_j(z)\equiv{\sum_{k\in V_n}}' \frac{\gamma_k
w_j^{(k)}}{z-\lambda_k},
\]
appearing in Eq.\ (\ref{chi1}). After some elementary
transformations the above sum reduces to
\[
\widetilde\Sigma_j(z)=\frac{\varepsilon}{n^{1+\delta}}\sum_{k_1=2}^n\frac{\cos\left[2\pi
(j_1-1)(k_1-1)/n\right]}{z-d+1-\cos\left[2\pi(k_1-1)/n\right]},
\]
and the summation technique from \cite{p94} yields
\[
\widetilde\Sigma_j(z)=\frac{2\varepsilon}{n^\delta (x_2-x_1)}
\frac{x_2^{j_1-1}+x_2^{n-j_1+1}}{x_2^n-1}-\frac{\varepsilon}{n^{1+\delta}(z-d)}.
\]

On application of the saddle-point method to the integral in Eq.\
(\ref{chi1}) one obtains the following expression for the characteristic
functions of random variables $x_j$ and $x_l$:
\begin{equation}
\chi(t,s)\sim\exp\left[-{\sum_{k\in V_n}}'
\frac{\left(tw_j^{(k)}+sw_l^{(k)}\right)^2}{4\beta
J\left(z^*_n-\lambda_k\right)}+
it\left(\widetilde\Sigma_j(z^*_n)+\rho\right)+
is\left(\widetilde\Sigma_l(z^*_n)+\rho\right)\right].
\label{chi3}
\end{equation}
Hence, as it is usually the case for pure states of the spherical
model, for large values of $n$ the random variables $\{x_j,j\in V_n\}$
have asymptotically normal distributions.

If $\delta\in(0,1)$, then the saddle point is given by
$z_n^*=\lambda_{\,\rm s}+\zeta^* n^{-2(1+2\delta)/3}$, see Eq.\
(\ref{sp2}), and the large-$n$ asymptotics of the expected values
of the microscopic variables (the multipliers of $it$ and $is$ in
Eq.\ (\ref{chi3})) are given by
\[
\langle x_j\rangle\sim \frac{\varepsilon
n^{(1-\delta)/3}}{2J\sqrt{2\zeta^*}}
\left[1+\sqrt{2\zeta^*}n^{-(1+2\delta)/3}\right]^{-j_1},
\]
if $j_1\leq n/2$. In this case the droplet shape is clearly
deformed by the external field, since virtually all ``spherical
matter" gathers in a narrow layer of the width $\sim
n^{(1+2\delta)/3}$ around the hyperplane $j_1=1$, where the field
is applied.

If $\delta=1$, then the saddle point is given by
$z_n^*=\lambda_{\,\rm s}+\zeta^* n^{-2}$, where $\zeta^*$ is a
solution of Eq.\ (\ref{sp3}), and the large-$n$ asymptotics of the
expected values of the microscopic variables are given by
\begin{equation}
\langle x_j\rangle\sim \rho+\frac{\varepsilon }{2J}\left[
\frac{\cosh\left[(1-2\gamma_1)\sqrt{\zeta^*/2-\pi^2}\right]}
{\sqrt{2(\zeta^*-2\pi^2)}\sinh\sqrt{\zeta^*/2-\pi^2}}-
\frac{1}{\zeta^*-2\pi^2}\right].
\label{evs2}
\end{equation}
if $j_1\sim\gamma_1 n$. In this case the droplet shape is still
deformed by the external field, since it depends on the field's
magnitude $\varepsilon$.

Finally, if $\delta\in(1,d-1)$, then the large-$n$ asymptotics of
the expected values of the thermodynamic variables are given by
\[
\langle x_j\rangle\sim
\rho+\sqrt{2(1-\rho^2)(1-\beta_c/\beta)}\cos\left(2\pi
\gamma_1\right).
\]
if $j_1\sim\gamma_1 n$. In this case, apparently, the droplet
shape is not deformed by the external field, since it does not
depend on the field's magnitude $\varepsilon$, and it is the same
as in Eqs.\ (\ref{shape1}) and (\ref{shape2}).

Thus, the droplet shape is sensitive to the magnitude of the
symmetry-breaking field only if it scales with $n$ as
$n^{-\delta}$ with $\delta\in[0,1]$. If the field is switched off
reasonably fast, $\delta\in(1,d-1)$ in Eq.\ (\ref{sbf}),
the droplet shape is not deformed by the field
(does not depend on the magnitude of the field). If the field is
switched off too fast, $\delta\geq d-1$, then it does not transfer
the thermodynamic system into a pure state.

It is appropriate to stress at this point that different
configurations of the symmetry breaking field $\{h_j: j\in V_n\}$
can produce different pure macroscopic phases. Therefore the
droplet shape is sensitive to the {\em type} of field used for
quasi-averaging. The eigenvector (\ref{qah}) is only one of $2d$
orthogonal vectors spanning the eigenspace corresponding to the
second-largest eigenvalue of the interaction matrix. In this
respect, the symmetry breaking field (\ref{sbf}) is special,
because it is orthogonal to all of the remaining $2d-1$
eigenvectors corresponding to the eigenvalue $\lambda_s$. If
instead of the field (\ref{sbf}) we decide to use a configuration
$\{h_j: j\in V_n\}$, which has a non-zero scalar product, say,
with
\[
\sqrt{\frac{2}{n}}
\cos\left[\frac{2\pi(j_2-1)}{n}-\alpha\right],\quad
j\equiv(j_1,j_2,\ldots,j_d)\in V_n,
\]
then the corresponding vector of expected values $\langle
x_j\rangle$, $j\in{V_n}$ will contain a component proportional to
$\cos(2\pi j_2/n)$, cf.\ Eq.\ (\ref{tocf}).

\section{Discussion.}

The investigation of droplet shape performed in this paper
demonstrates the importance of decomposing mixed
states into constituent pure components. It appears that, often, mixed
states are mathematical constructions reflecting absence of sufficient
information about the system under investigation. A priori we do not know
which of many possible pure states of the system will be actually observed
in an experiment. As a consequence, a direct mathematical solution of the
model produces a mixed state. At the same time, if an experimentalist actually
performs measurement in the corresponding physical system, the obtained
results would be described by one of pure states.

Whether this point of view is valid or not, depends, of course, on
how quickly the system under investigation can transit from one pure state into
another. Namely, if the typical transition time is much
larger than the typical experimental observation time, then measurements
yield results corresponding to one of pure states. If the observation
time is much longer than the transition time, then the observed
values are likely to correspond to a mixed state.

Investigation of dynamical properties of the mean spherical model
were performed recently, see \cite{gl00}. It is possible to calculate the
interstate transition times using the technique proposed there.
We have not attempted yet those calculations, and we do not even know
if the droplet shape in the mean spherical and the conventional spherical
model are similar. But, it does not seem inconceivable, that the transition
time from one pure state into another is much greater than the typical
measurement time. Indeed, investigations of large deviation probabilities
show that in order to transit from one pure state into another within the
conventional spherical model one has to overcome a free-energy
barrier that grows exponentially with $n^{d-2}$.

It is well known that gaussian approximations correspond to
leading orders of low-temperature expansions of various continuous
models. Therefore, one can hope that the behaviour of realistic
continuous models, for instance $O(n)$ models, is qualitatively
similar to the findings of the present paper. On the other hand, the
common knowledge is that physical substances in liquid phases form
droplets which boundaries are sharp in the macroscopic scale. Most
likely, the sharpness of boundaries should be attributed to the
atomistic (discrete) microscopic structure of physical substances. Therefore,
in this respect, discrete lattice gas models of Ising type provide much more
realistic description of various substances in the liquid state.


\begin{thebibliography}{99}
\bibitem{b60} N.\ N.\ Bogolubov, On some problems of the theory of superconductivity,
{\em Physica\/},\ Suppl.\ to vol.\ {\bf 26}:S1--S16 (1960).
\bibitem{az85} N.\ Angelescu and V.\ A.\ Zagrebnov, A generalized
quasiaverage approach to the description of the limit states of
the $n$-vector Curie-Weiss ferromagnet, {\em J.\ Stat.\ Phys.}\
{\bf 41}:323--334 (1985).
\bibitem{bzt86} J.\ G.\ Brankov, V.\ A.\ Zagrebnov, and
N.\ S.\ Tonchev, Description of the limit Gibbs states for the
Curie-Weiss-Ising model, {\em Theor.\ Math.\ Phys.}\ {\bf
66}:72--80 (1986).
\bibitem{dks92} R.\ L.\ Dobrushin, R.\ Koteck\'y, and S.\
Shlosman, {\em Wulff Construction: A Global Shape from Local
Interaction,\/} (AMS translation series, Providence, 1992).
\bibitem{gl00} C.\ Godr\`eche and J.\ M.\ Luck, Response of Non-Equilibrium
Systems at Criticality: Ferromagnetic Models in Dimension Two and Above,
{\em J.\ Phys.\ A: Math.\ Gen.}\ {\bf 33}:9141--9164 (2000).
\bibitem{gb62} H.\ A.\ Gersch and T.\ H.\ Berlin, Spherical Lattice Gas,
{\em Phys.\ Rev.}\ {\bf 127}:2276--2283 (1962).
\bibitem{gj81} J.\ Glimm and A.\ Jaffe,
{\em Quantum Physics: A Functional Integral Point of View.\/}
(Springer-Verlag, Berlin, 1981).
\bibitem{ms67} R.\ A.\ Minlos and Y.\ G.\ Sinai, The phenomenon of "phase
separation" at low temperatures in some lattice models of a gas. I,
{\em Matem.\ Sbornik.}\ {\bf
73}:375--448 (1967).
\bibitem{ms68} R.\ A.\ Minlos and Y.\ G.\ Sinai, The phenomenon of "phase
separation" at low temperatures in some lattice models of a gas. II,
{\em Trans. Moscow Math. Soc.}\ {\bf 19}:113--178 (1968).
\bibitem{p94} A.\ E.\ Patrick, The influence of external boundary conditions
on the spherical model of a ferromagnet. I. Magnetization
profiles, {\em J.\ Stat.\ Phys.}\ {\bf 75}:253--295 (1994).
\bibitem{pk60} W.\ Pressman and J.\ B.\ Keller,
Equation of State and Phase Transition of the Spherical Lattice
Gas, {\em Phys.\ Rev.}\ {\bf 120}:22--32 (1960).
\bibitem{s91} Ya.\ G.\ Sinai,
{\em Mathematical Problems in Statistical Mechanics,\/} (World-Scientific,
Singapore, 1991).
\end{thebibliography}
\end{document}